\documentclass[aps,prl,reprint,superscriptaddress]{revtex4-1}
\usepackage{physics}
\usepackage[pdftex]{graphicx}    
\graphicspath{{Figures/}}
\DeclareGraphicsExtensions{.pdf,.jpeg,.png,.jpg}   

\begin{document}
\title{Spin lifetime and charge noise in hot silicon quantum dot qubits}

\author{L. Petit}
\affiliation{QuTech and Kavli Institute of Nanoscience, TU Delft, P.O. Box 5046, 2600 GA Delft, The Netherlands}

\author{J. M. Boter}
\affiliation{QuTech and Kavli Institute of Nanoscience, TU Delft, P.O. Box 5046, 2600 GA Delft, The Netherlands}

\author{H. G. J. Eenink}
\affiliation{QuTech and Kavli Institute of Nanoscience, TU Delft, P.O. Box 5046, 2600 GA Delft, The Netherlands}

\author{G. Droulers}
\affiliation{QuTech and Kavli Institute of Nanoscience, TU Delft, P.O. Box 5046, 2600 GA Delft, The Netherlands}

\author{M. L. V. Tagliaferri}
\affiliation{QuTech and Kavli Institute of Nanoscience, TU Delft, P.O. Box 5046, 2600 GA Delft, The Netherlands}

\author{R. Li}
\affiliation{QuTech and Kavli Institute of Nanoscience, TU Delft, P.O. Box 5046, 2600 GA Delft, The Netherlands}

\author{D. P. Franke}
\affiliation{QuTech and Kavli Institute of Nanoscience, TU Delft, P.O. Box 5046, 2600 GA Delft, The Netherlands}

\author{K. J. Singh}
\affiliation{Components Research, Intel Corporation, 2501 NE Century Blvd, Hillsboro, OR 97124, USA}

\author{J. S. Clarke}
\affiliation{Components Research, Intel Corporation, 2501 NE Century Blvd, Hillsboro, OR 97124, USA}

\author{R. N. Schouten}
\affiliation{QuTech and Kavli Institute of Nanoscience, TU Delft, P.O. Box 5046, 2600 GA Delft, The Netherlands}

\author{V. V. Dobrovitski}
\affiliation{QuTech and Kavli Institute of Nanoscience, TU Delft, P.O. Box 5046, 2600 GA Delft, The Netherlands}

\author{L. M. K. Vandersypen}
\affiliation{QuTech and Kavli Institute of Nanoscience, TU Delft, P.O. Box 5046, 2600 GA Delft, The Netherlands}

\author{M. Veldhorst}
\affiliation{QuTech and Kavli Institute of Nanoscience, TU Delft, P.O. Box 5046, 2600 GA Delft, The Netherlands}

\begin{abstract}
We investigate the magnetic field and temperature dependence of the single-electron spin lifetime in silicon quantum dots and find a lifetime of 2.8 ms at a temperature of 1.1 K. We develop a model based on spin-valley mixing and find that Johnson noise and two-phonon processes limit relaxation at low and high temperature respectively. We also investigate the effect of temperature on charge noise and find a linear dependence up to 4 K. These results contribute to the understanding of relaxation in silicon quantum dots and are promising for qubit operation at elevated temperatures.
\end{abstract}

\maketitle

Electron spins in semiconductor quantum dots \cite{loss1998quantum} are considered to be one of the most promising platforms for large-scale quantum computation. Silicon can provide key assets for quantum information processing, including long coherence times \cite{veldhorst2014addressable, yoneda2017quantum},  high-fidelity single-qubit rotations  \cite{veldhorst2014addressable, yoneda2017quantum} and two-qubit gates \cite{veldhorst2015two, zajac2018resonantly, watson2017programmable}, which have already enabled the demonstration of quantum algorithms \cite{watson2017programmable}. Quantum dots based on silicon metal-oxide semiconductor (Si-MOS) technology provide additional prospects for scalability due to their compatibility with conventional manufacturing technology \cite{li2017crossbar, taylor2005fault}, which opens the possibility to co-integrate classical electronics and qubits on the same wafer to avoid an interconnect bottleneck \cite{veldhorst2017silicon, vandersypen2017interfacing}. However, control electronics will introduce a power dissipation that seems incompatible with the available thermal budget at temperatures below 100 mK, where qubits currently operate. Understanding and improving the robustness of qubits against thermal noise is therefore crucial, while operating qubits beyond 1 K could entirely resolve this challenge.

Spin relaxation and charge noise are two essential metrics for quantum dot qubits. While the spin lifetime $T_1$ can be of the order of seconds in silicon quantum dots \cite{xiao2010measurement, simmons2011tunable, Yang2013spin}, exceeding by orders of magnitude the dephasing time $T_2^*$ \cite{veldhorst2014addressable}, it is presently unclear how $T_1$ will be affected by temperature and whether it will become the shortest timescale for quantum operations at elevated temperatures. Spin qubits are also sensitive to charge noise, and electrical fluctuations can reduce qubit readout and control fidelities. The temperature dependence of these two parameters is therefore vital in evaluating the prospects for hot spin qubits.  

\begin{figure}[!h]
	\centering
	\includegraphics[width=1\linewidth]{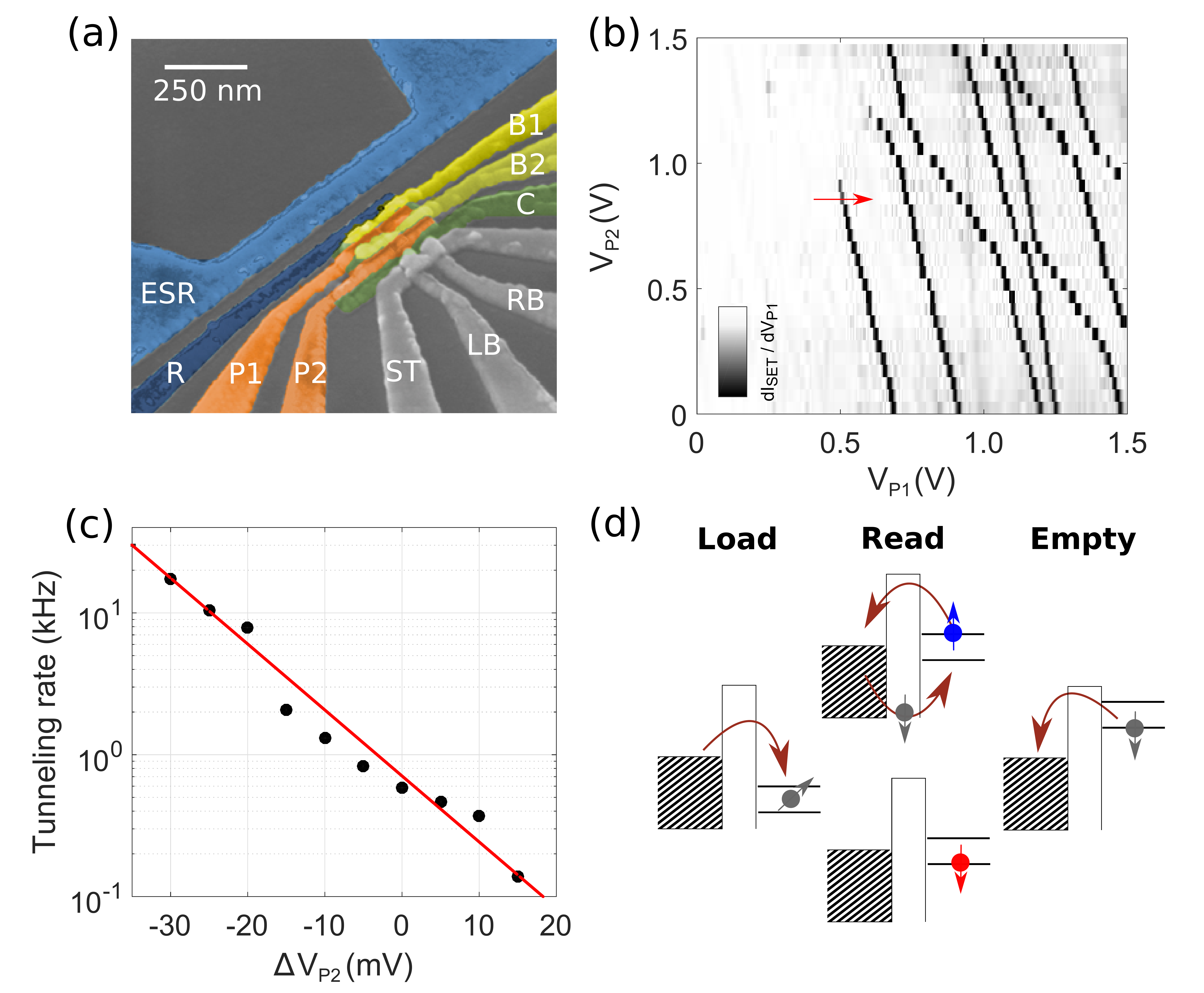}
	\caption{(a) Scanning electron microscope image of a device identical to the one measured. R is the reservoir gate, P1, P2, B1 and B2 are the plunger gates, and C confines the electrons in the dots. LB and RB are the left and right barrier of the quantum dot used for sensing, and ST is used both as top gate and reservoir. The ESR line can be used for spin manipulation. (b) Charge stability diagram of the device measured via a double lock-in technique \cite{yang2011dynamically} (see Supplemental Material \cite{SuppInfo}). The transition lines, due to the different slope, can be attributed to three coupled quantum dots. The red arrow shows the (0$\rightarrow$1) charge transition relevant for the experiment. (c) Tunneling rate between the dot and the reservoir as a function of V\textsubscript{P2}. $\Delta$V\textsubscript{P2}=0 corresponds to the value set during the experiment. The red line is an exponential fit. (d) Pulsing sequence used to perform single-shot readout of the electron spin \cite{elzerman2004single} in the case $E_z < E_{vs}$. Above the valley splitting there is also an intermediate level between the ground and excited spin state, corresponding to the spin-down state of the excited valley.}
	\label{Fig:Device}
\end{figure}

Here we investigate in detail the temperature dependence of spin relaxation  and charge noise of a Si-MOS quantum dot. We construct a model based on direct and two-phonon transitions including all spin and valley states of the lowest orbital. The model provides good agreement with the experiments and we conclude that while at low temperatures $T_1$ is limited by Johnson noise, probably originating from the two-dimensional electron gas (2DEG) channels present in the device, two-phonon processes determine the relaxation rate above $200$ mK. Based on our results we predict how the spin lifetime can be improved by decreasing the magnetic field and increasing the valley-splitting energy. Furthermore, we investigate the charge noise and measure a rather weak temperature dependence.

Figure \ref{Fig:Device} (a) shows a scanning electron microscope (SEM) image of a quantum dot device, realized in isotopically enriched silicon (\textsuperscript{28}Si), identical in design to the one measured. Figure \ref{Fig:Device} (b) presents the charge stability diagram of the device, showing charge transitions originating from three quantum dots, and we deplete one quantum dot to the last electron. From the temperature dependence of the transition width (see Supplemental Material \cite{SuppInfo}) we extract a lever arm $\alpha_{P1}= 0.12$ eV/V. We tune the tunnel rate between the quantum dot and the reservoir by controlling the gate P2 (see Fig.~\ref{Fig:Device} (c)), which moves the position of the quantum dot thereby changing the distance to the reservoir. During the experiment, since the DC signal of the sensing dot is filtered with a 2 kHz low pass filter, the dot-reservoir tunnel rate is set to approximately 700 Hz. 

As shown in Fig.~\ref{Fig:Device} (d), we measure the spin lifetime by applying a three-level voltage pulse to the gate P1, while monitoring the DC current of the sensing dot. First, we inject an electron into the quantum dot, we read out the spin state, and we finally empty the quantum dot \cite{elzerman2004single}. An additional level is added to the pulse after the empty phase in order to cancel out any DC offset. We measure the spin-up fraction as a function of load time and extract $T_1$ by fitting the data with an exponential decay.

\begin{figure}[]
	\centering
	\includegraphics[width=1\linewidth]{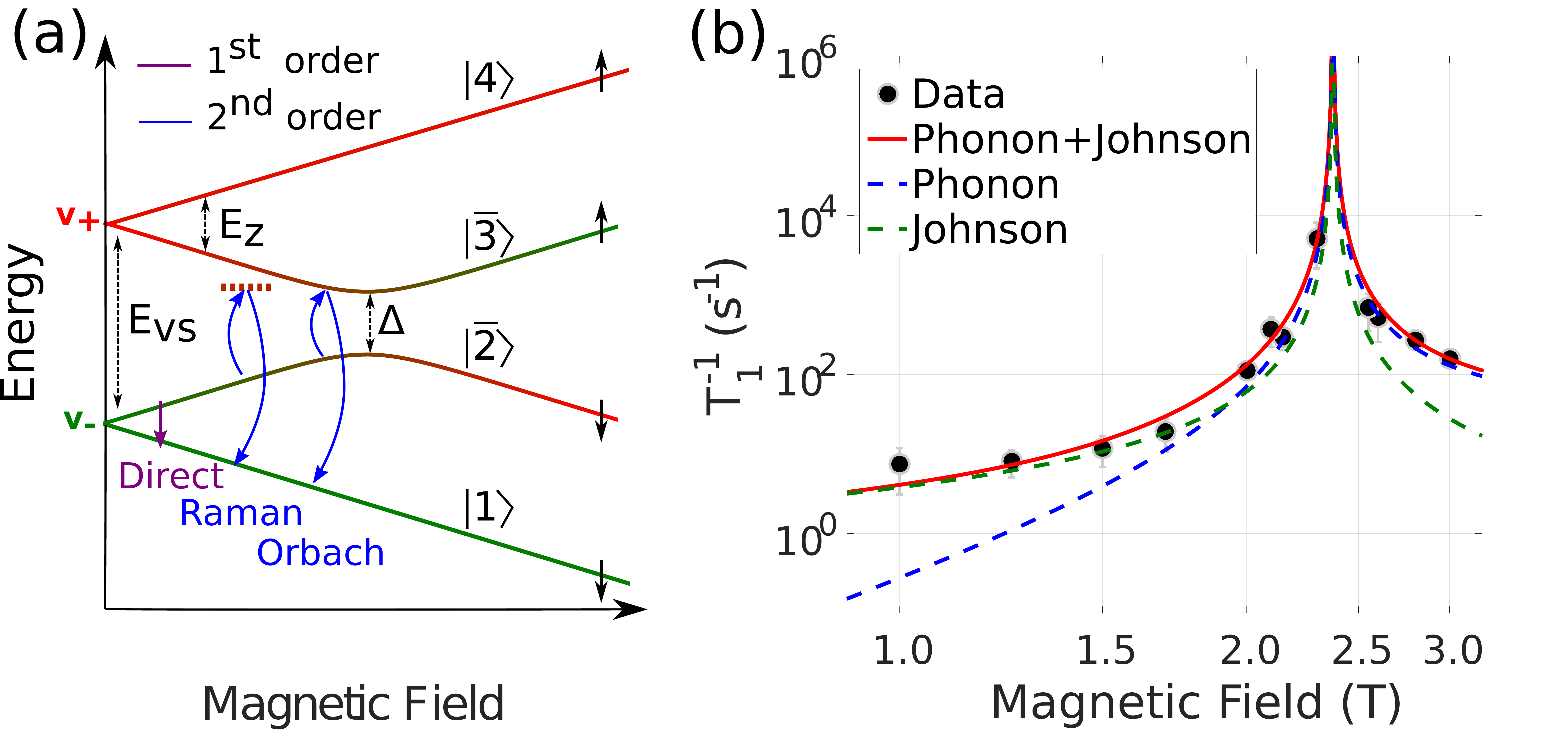}
	\caption{(a) Energy levels in a silicon quantum dot, showing both valley and spin degrees of freedom. As an example, the transition $\Gamma_{\bar{2}\bar{1}}$ is sketched in first-order and in second-order via virtual and resonant transitions. (b) Relaxation rate as a function of magnetic field. The fittings include contributions from Johnson and phonon mediated relaxation obtained through the model explained in the main text. From the fittings of the magnetic field and temperature dependence we extract $E_{vs}=275$ $\mu$eV, $\Gamma_0	^J(E_{vs}/\hbar)=2 \cdot 10^{-12}$ s, $\Gamma_0	^{ph}(E_{vs}/\hbar)=6 \cdot 10^{-12}$ s and $\Delta=0.4$ neV.}
	\label{Fig:Relaxation}
\end{figure}

\begin{figure}
	\centering
	\includegraphics[width=1\linewidth]{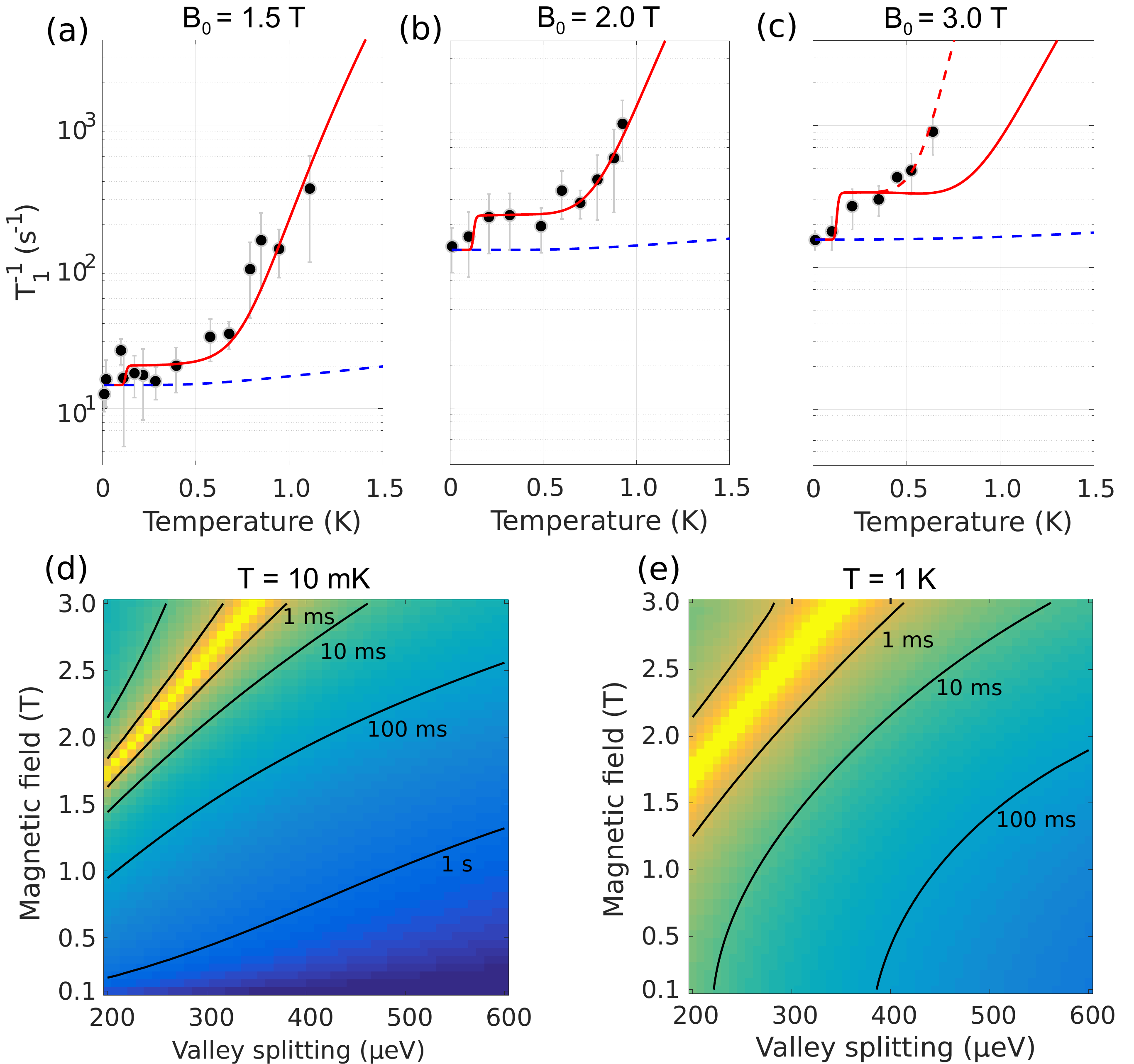}
	\caption{(a-c) Temperature dependence of the relaxation rate at $B_0=1.5$ T (a), 2 T (b) and 3 T (c). The red line is a fit taking into account Johnson and phonon noise in first and second-order. The red dashed line includes possible contributions coming from the coupling with the excited orbital states. First-order processes are shown in the dashed blue line. (d,e) Relaxation rate as a function of magnetic field and valley splitting for $T=10$ mK (d) and for $T=1$ K (e) as extracted from the model discussed in the main text.}
	\label{Fig:TemperatureDependence}
\end{figure}

The measured $T_1$ as a function of magnetic field (applied in the [010] direction) is plotted in Fig.~\ref{Fig:Relaxation} (b) and the temperature dependence for three different magnetic fields is shown in Fig.~\ref{Fig:TemperatureDependence} (a), (b) and (c). Thermal broadening of the reservoir limits the experimentally accessible regime. At base temperature (fridge temperature $< 10$ mK, electron temperature 108 mK, see Supplemental Material \cite{SuppInfo})  we measure a maximum $T_1$ of 145 ms at $B_0=1$ T. We find that even when increasing the temperature to 1.1 K, $T_1$ is 2.8 ms. This is more than an order of magnitude larger than the longest $T^*_2$ reported in silicon quantum dots \cite{veldhorst2014addressable}.  

In order to understand the magnetic field and temperature dependence of the relaxation rate, we need to consider the mixing between spin and valley. In silicon the four lowest spin-valley states are \cite{zwanenburg2013silicon}: $\ket{1}=\ket{v_{-},\downarrow}$, $\ket{2}=\ket{v_{-},\uparrow}$, $\ket{3}=\ket{v_{+},\downarrow}$, $\ket{4}=\ket{v_{+},\uparrow}$ (see Fig.~\ref{Fig:Relaxation} (a)). In presence of interface disorder, spin-orbit interaction can couple states with different valleys and spins, introducing a channel for spin relaxation \cite{Yang2013spin}. This leads to the eigenstates $\ket{1}, \ket{\bar{2}}, \ket{\bar{3}}, \ket{4}$, where:
\begin{eqnarray}
\ket{\bar{2}} = \bigg(\frac{1-a}{2}\bigg)^{1/2} \ket{2} -  \bigg(\frac{1+a}{2}\bigg)^{1/2} \ket{3}  	
\label{eq:state2}
\\
\ket{\bar{3}} =  \bigg(\frac{1+a}{2}\bigg)^{1/2} \ket{2} + \bigg(\frac{1-a}{2}\bigg)^{1/2} \ket{3} .
\label{eq:state3}
\end{eqnarray}
Here we have $a = - (E_{vs} - \hbar \omega_{z}) / \sqrt{(E_{vs} - \hbar \omega_{z})^2 + \Delta ^2}$, where $\Delta$ is the splitting at the anticrossing point of the states $\ket{2}$ and $\ket{3}$, $E_{vs}$ is the valley splitting and $\hbar \omega_{z}$ the Zeeman energy. In the presence of electric fields, the electrons in the excited states $\ket{\bar{2}}$ and $\ket{\bar{3}}$ can relax to the ground state $\ket{1}$, because they are in an admixture of spin and valley states. We define a relaxation rate $\Gamma_{sv}$, corresponding to $\Gamma_{\bar{2}1}$  and $\Gamma_{\bar{3}1}$ before and after the anticrossing, respectively. The resulting expression is \cite{huang2014spin}:
\begin{eqnarray}
\Gamma_{sv} = \Gamma_{v_{+}v_{-}}(\omega_z) F_{sv}(\omega_z) 
\label{eq:spinvalleyRel}
\end{eqnarray}
where $\Gamma_{v_{+}v_{-}}$ is the pure valley relaxation rate and $F_{sv} (\omega_z) = (1-\abs{a(\omega_z)})$. When $E_{vs}=E_z$, the function $F_{sv}$ peaks and the spin relaxation equals the fast pure valley relaxation \cite{Yang2013spin}. From the location of this relaxation hot spot we determine a valley splitting $E_{vs}$ of 275 $\mu$eV, comparable with values reported in other works \cite{veldhorst2014addressable}.

\begin{figure}
	\centering
	\includegraphics[width=1\linewidth]{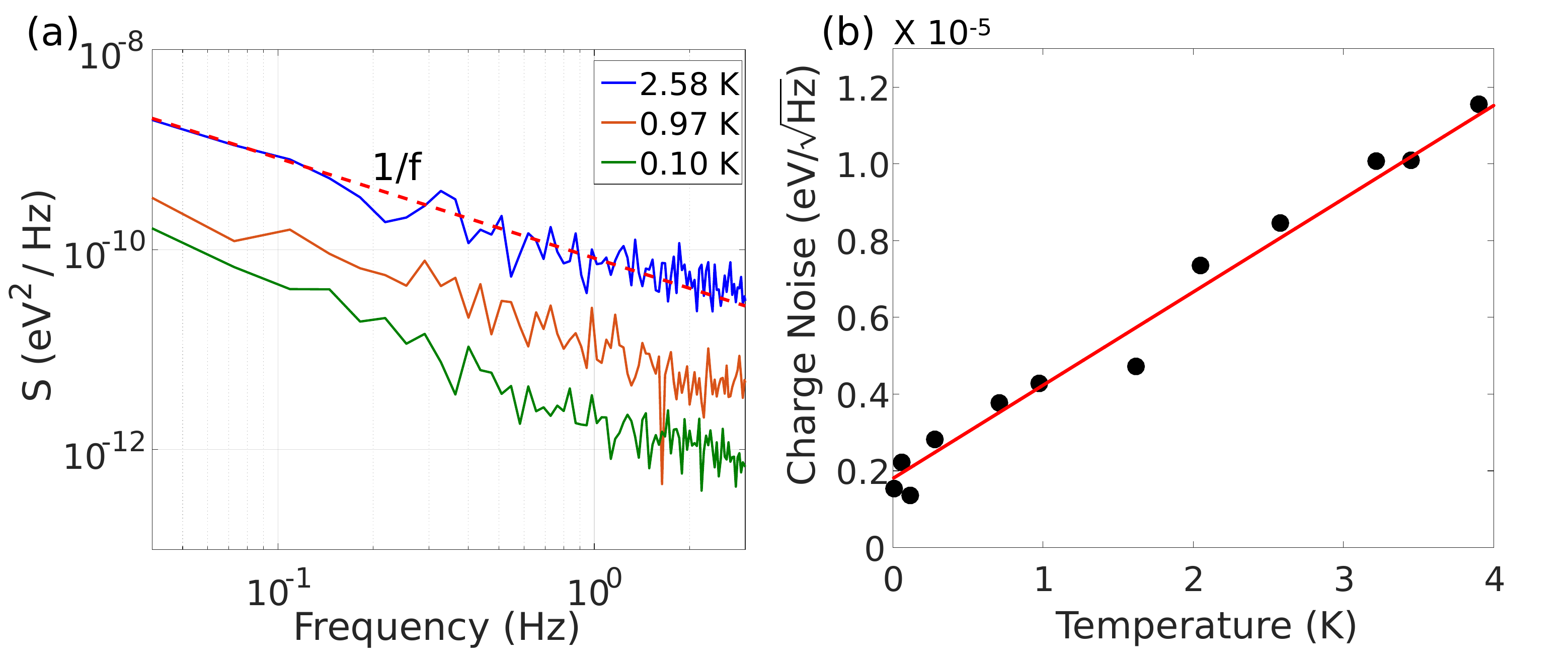}
	\caption{(a) Charge noise spectra obtained for three different temperatures. At higher frequencies the 1/f signal is masked by white noise. (b) Charge noise at a frequency of 1 Hz as a function of temperature fitted with a linear function.}
	\label{fig:ChargeNoise}
\end{figure}

Possible sources of electrical noise include $1/f$ charge noise, Johnson noise and phonon noise. We measure small values for charge noise (see Fig.~\ref{fig:ChargeNoise}) and thus neglect their contribution, further justified by the high frequencies of 20-100 GHz, associated with the Zeeman energies studied here ($1~\textrm{T}<B_0<3~\textrm{T}$). We also neglect the Johnson noise coming from the circuits outside the dilution refrigerator since all room temperature electronics are well filtered. The most relevant of these noise sources is the arbitrary waveform generator used to apply voltage pulses. However, the corresponding lines are attenuated by 12 dB and have an intrinsic cut-off frequency of 1 GHz, making the noise in the 20-100 GHz range negligible. Another possible source of Johnson noise is the resistive 2DEG, which generates electric field fluctuations that have a capacitive coupling to the quantum dot. In the present device, the main contribution is likely due to the 2DEG underneath the reservoir gate, which is in close proximity to the quantum dot.   
 
The remaining contributions are Johnson noise and phonons. The pure valley relaxation for these two cases is given by \cite{Yang2013spin, huang2014spin}:     
\begin{eqnarray}
\Gamma_{v_{+}v_{-}}^J (\omega) = \Gamma_0	^J \cdot \bigg(\frac{\omega}{\omega_{vs}}\bigg) [1+2n_b(\hbar \omega, k_{\textrm{B}}T)]
\label{eq:johnsonSpec}
\\
\Gamma_{v_{+}v_{-}}^{ph} (\omega) =  \Gamma_0	^{ph} \cdot \bigg(\frac{\omega}{\omega_{vs}}\bigg)^5 [1+2n_b(\hbar \omega, k_{\textrm{B}}T)] ,
\label{eq:phononSpec}
\end{eqnarray}
where $\hbar \omega$ is the energy difference, $\omega_{vs} = E_{vs}/\hbar$ is a normalization constant and $n_b$ is the Bose-Einstein distribution. The two contributions can be distinguished by the different magnetic field dependence that follows from $\omega_z F_{sv}(\omega_z)$ in the case of Johnson noise and from $\omega_z^5 F_{sv}(\omega_z)$ for phonons. As shown in Fig.~\ref{Fig:Relaxation} (b) the magnetic field dependence of $T_1$ at base electron temperature can be explained in terms of Johnson mediated relaxation dominant at low fields, and a phonon contribution, mainly relevant for $\hbar \omega_z > E_{vs}$.

We now turn to the temperature dependence, shown in Fig.~\ref{Fig:TemperatureDependence} (a), (b) and (c). As shown in Eq. (\ref{eq:johnsonSpec}) and (\ref{eq:phononSpec}), the temperature dependence is the same to first-order for phonon and Johnson noise and given by $1+2n_b(\hbar\omega_z, k_{\textrm{B}}T)$. 
If $\hbar\omega_z \gg k_\textrm{B} T$ spontaneous phonon emission dominates and the relaxation rate is temperature independent, while for $\hbar\omega_z \ll k_\textrm{B}T$ it increases linearly with temperature. The relaxation rates caused by first-order processes are shown by the blue lines in Fig.~\ref{Fig:TemperatureDependence} (a), (b) and (c), which fit the low temperature region of the plots. However, the same processes cannot justify the rapid increase of $T_1$ measured at higher temperatures. In order to explain the full temperature dependence we also need to take into account two-phonon processes.

As depicted in Fig.~\ref{Fig:Relaxation} (a), these transitions happen in a two-step process via intermediate states. These intermediate transitions can be energy-conserving and energy non-conserving (virtual) processes, since energy must be conserved only between the initial and the final state. We obtain a two-phonon process by expanding the spin-phonon interaction in second-order perturbation theory \cite{shrivastava1983theory}: 
\begin{equation}
\Gamma^{(2)}_{if} =\frac{2 \pi}{\hbar} \abs{\sum_{k} \frac{V_{fk}V_{ki}}{E_i - E_k + \frac{1}{2} i \hbar \Gamma_k}}^2 \delta(E_i - E_f) ,
\label{eq:2nd_order}
\end{equation}

where $V_{fk}$, $V_{ki}$ are the matrix elements between the states and $1/\Gamma_k$ is the lifetime of the intermediate state, which depends on all first-order processes between $k$ and the other states. The square of the matrix elements is proportional to the valley relaxation rate $\Gamma_{v_+v_-}$. Relaxation through Johnson noise can also be expanded in second-order perturbation theory, however the temperature dependence is much weaker (see Supplemental Material \cite{SuppInfo}) and its contribution will therefore be neglected.     

Since the thermal energy is comparable to the level splitting in the temperature window 0.5-1 K, absorption processes cannot be neglected. In order to understand the relaxation dynamics we have developed a model that includes all possible transitions between the four spin-valley states in first and second-order. For completeness, we have also included in the model the weak coupling between the states $\ket{1}$ and $\ket{4}$. We evaluate all the transition rates and we use them to solve a 4x4 system of coupled differential rate equations given by:
\begin{equation}
\frac{dN_i}{dt} = -N_i \sum_{j \neq i} \Gamma_{ij} +  \sum_{j \neq i} \Gamma_{ji} N_j \;\;\; \text{for} \;\;\; i,j=1,\bar{2},\bar{3},4,
\label{eq:Rates}
\end{equation}
$N_i$ being the population of the state $i$. The red lines in Fig.~\ref{Fig:TemperatureDependence} (a), (b) and (c), show the relaxation rates as obtained from Eq. \ref{eq:spinvalleyRel}, 
\ref{eq:2nd_order} and \ref{eq:Rates} (see also Supplemental Material \cite{SuppInfo}). The good agreement between model and experiment provides an indication that, even at high temperatures, relaxation is dominated by spin-valley physics. The rates relevant to the relaxation process are found to be the spin-flip transitions involving the three lowest states: $\Gamma_{\bar{2}, 1}$, $\Gamma_{\bar{2}, \bar{3}}$ and $\Gamma_{\bar{3}, 1}$, $\Gamma_{\bar{3}, \bar{2}}$ at $E_z$ below and above $E_{vs}$ respectively.
The relaxation rate above 200 mK consists of a flat region followed by a rising part. We attribute this behavior to the second-order process described by Eq. \ref{eq:2nd_order}. We consider separately the contributions of the resonant ($\abs{E_i - E_k} \ll \hbar \Gamma_k$) and off-resonant transitions ($\abs{E_i - E_k} \gg \hbar \Gamma_k$). In the first case, known as Orbach process \cite{orbach1961spin}, the second-order relaxation is proportional to $\abs{V_{fk}V_{ki}}^2/\Gamma_k$ (see Supplemental Material \cite{SuppInfo}). At sufficiently low temperatures, the spin lifetime depends exponentially on the temperature since the numerator is proportional to $n_b$ and the denominator is temperature independent. We therefore theoretically predict the brief steep rise around 150-200 mK. At high temperatures $\Gamma_k$ also becomes proportional to $n_b$ and the temperature dependence vanishes. This explains the main flat region that we observe in Fig. \ref{Fig:TemperatureDependence} (a), (b) and (c). For off-resonant transitions, known as Raman process, the relaxation rate scales polynomially with the temperature. As discussed in the Supplemental Material \cite{SuppInfo}, in case of phonon-mediated transitions, a $T^9$ temperature dependence is obtained. The Raman process dominates over the Orbach process above 500 mK (see Fig.~\ref{Fig:TemperatureDependence} (a), (b) and (c)). 

As we can see from Fig.~\ref{Fig:TemperatureDependence} (c), the increase in the relaxation rate at $B_0=3$ T does not match the model predictions above 500 mK, suggesting contributions to the relaxation from a different source rather than the valley mixing. We rule out second-order contributions from Johnson noise because of the much weaker temperature dependence. Possible contributions might come from a second-order process involving the excited orbital states, which is expected to give a $T^{11}$ temperature dependence as discussed in the Supplemental Material \cite{SuppInfo}. Coupling to orbital states can potentially give a magnetic field dependence that would make it not observable at lower fields. Coupling to orbital states mediated by direct processes give rise to a $B_0^2$ field dependence; this phenomenon is known as Van Vleck cancellation, a consequence of Kramer's theorem \cite{abragam2012electron}. For two-phonon processes, Van Vleck cancellation together with the spin-valley mixing can potentially give an even stronger field dependence. 

The spin lifetime can be increased by reducing the spin-valley coupling. As shown in Eq. \ref{eq:state2} and \ref{eq:state3}, it can be strongly increased by reducing the applied magnetic field or by increasing the valley splitting energy. In Si-MOS the valley splitting can be electrically controlled and increased to $E_{vs}\approx1$ meV \cite{veldhorst2014addressable, laucht2015electrically}. Figure \ref{Fig:TemperatureDependence} (d) and (e) show the magnetic field and the valley splitting energy dependence of the relaxation rate for $T=10$ mK and $T=1$ K, using the parameters extracted from our numerical fittings of the experimental data. These results predict a spin lifetime at 1 K of approximately 500 ms, when $B_0=0.1$ T and $E_{vs}=575$ $\mu$eV. The relaxation at low magnetic fields is predicted to be dominated by second-order processes even at low temperature, due to the stronger field dependence of the first-order processes.   

We now turn to charge noise measurements. In a minimal model, charge noise can be attributed to defects that can trap or release charges, giving rise to electrical noise with a characteristic 1/f spectrum \cite{paladino20141}. We measure the charge noise in our device as current fluctuations of the sensing dot tuned to a regime with a high slope $dI/dV$, to maximize the sensitivity. The time trace of the current is converted to voltage noise by dividing by the slope; then the spectrum is obtained through a Fourier transform. The same process is repeated in Coulomb blockade in order to subtract the baseline noise coming from the electronics \cite{freeman2016comparison}. Finally, the voltage fluctuations are converted to energy fluctuations by using the lever arm $\alpha_{ST}=0.18$ eV/V of the sensing dot. The spectra shown in Fig.~\ref{fig:ChargeNoise} (a), scale as 1/f for the probed frequency regime. Fig.~\ref{fig:ChargeNoise} (b) shows the temperature dependence of the charge noise at a fixed frequency of 1 Hz. We observe a linear increase of the charge noise over more than one decade of temperature (0.1-4 K), changing from approximately 2 $\mu \textrm{eV}/\sqrt{\textrm{Hz}}$ to 12 $\mu \textrm{eV}/\sqrt{\textrm{Hz}}$. This is indicating a different relation than predicted by a simple model, which assumes an equal distribution of thermally activated fluctuators with relaxation rates distributed according to a Lorentzian. This model would give rise to a square root temperature dependence of the charge noise amplitude \cite{paladino20141}. The offset measured at low temperature can be attributed to electrical noise that couples to the sensing dot via the gates. This remarkably weak dependence suggests that qubit operation will only be moderately affected when increasing temperature.

In summary, we have investigated the magnetic field and temperature dependence of the spin lifetime and measured $T_1=2.8$ ms at 1.1 K and $T_1=145$ ms at base temperature. Relaxation occurs through electric field fluctuations that cause spin transitions mediated by spin-valley coupling. At temperatures below 200 mK the dominant noise source is Johnson noise, while second-order phonon processes dominate at higher temperatures. We have also shown how the spin lifetime can be further improved by operating in low magnetic fields and tuning to high valley splitting energies. In particular Si-MOS devices have the advantage of a large and tunable valley splitting, whereas in Si/SiGe it is typically not larger than $100$ $\mu$eV \cite{kawakami2014electrical}. Future work aimed at improving lifetimes could focus on schemes that do not explicitly require a large magnetic field, such as readout via Pauli spin blockade. In addition, we have measured the temperature dependence of the charge noise and find consistency with a linear trend from 100 mK to 4 K.

Leading solid-state approaches for large-scale quantum computation focus on decreasing the operation temperature down to the milliKelvin regime. Instead, the long spin lifetimes at elevated temperatures and the weak charge noise reported here indicate that such low temperatures are not a fundamental requirement for spins in Si-MOS quantum dots, providing an avenue for the demonstration of spin qubits with operation temperatures above one Kelvin.

\begin{acknowledgments}
M. V.  acknowledges  support  by  the  Netherlands Organization of Scientific Research (NWO) VIDI program.
Research was sponsored by the Army Research Office (ARO) and was accomplished under Grant No. W911NF-
17-1-0274. The views and conclusions contained in this document are those of the authors and should not be
interpreted as representing the official policies, either expressed or implied, of the Army Research Office (ARO), or the U.S. Government. The U.S. Government is authorized to reproduce and distribute reprints for Government purposes notwithstanding any copyright notation herein.
\end{acknowledgments}

\providecommand{\noopsort}[1]{}\providecommand{\singleletter}[1]{#1}%

\clearpage
\onecolumngrid
\appendix*
\begin{center}
	\textbf{\large Supplemental Material}
\end{center}

\section{Section 1: Device fabrication}
Fabrication starts from a 300 mm silicon wafer, upon which a 100 nm layer of epitaxial \textsuperscript{28}Si is grown, with a residual concentration of \textsuperscript{29}Si at 900ppm. The wafer is then thermally oxidized to form a 10 nm SiO\textsubscript{2} layer on top. We deposit 10 nm of Al\textsubscript{2}O\textsubscript{3}, fabricated by atomic layer deposition, and we pattern the nanostructures via electron-beam lithography and deposit three layers of Ti (3 nm) / Pd (37 nm) that are electrically isolated by Al\textsubscript{2}O\textsubscript{3} (7 nm). The smaller grain size of palladium films, compared to that of aluminum \cite{angus2007gate}, enables a more uniform set of electric gates \cite{brauns2018palladium}.

\section{Section 2: Rate equations}
\begin{figure}[]
	\centering
	\includegraphics[width=1\linewidth]{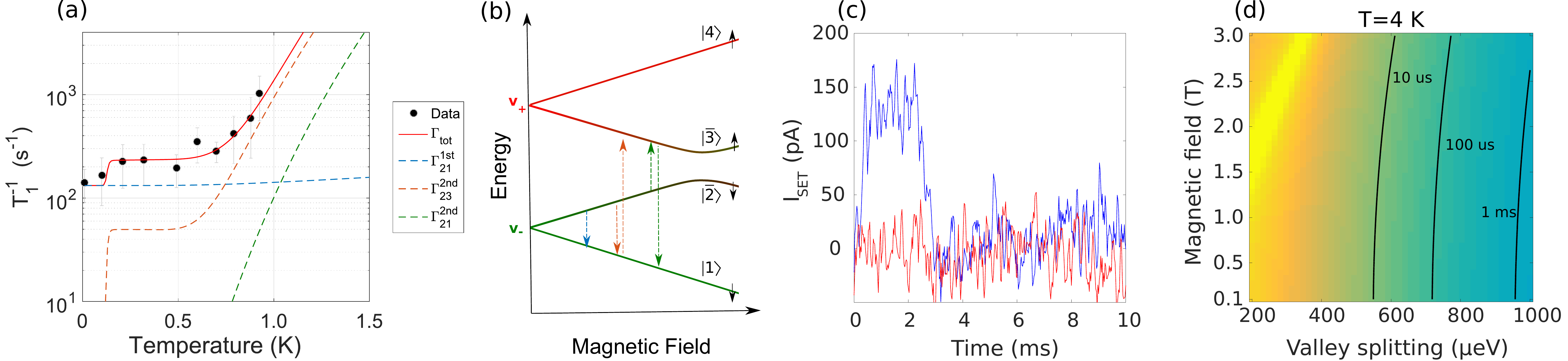}
	\caption{(a) Relaxation rate as a function of temperature for $B_0$=2 T. The dashed coloured lines show the relevant transition rates, including the first-order process between the states $\ket{\bar{2}}$ and $\ket{1}$ and second-order transitions via the states $\ket{1}$ and $\ket{\bar{3}}$. (b) The relevant transitions shown in the plot in (a) are sketched in an energy diagram. (c) Example of a single-shot readout trace. The blue curve is the response of the SET when a spin-up electron is readout. The red line is the analogous case for a spin-down electron. (d) Magnetic field and valley splitting dependence of the relaxation rate at a temperature of 4 K. Lifetimes larger than 1 ms are accessible with a valley splitting close to 1 meV. } 
	\label{Fig:AllRates}
\end{figure}

Figure \ref{Fig:AllRates} (a) and (b) shows the relevant contributions to the relaxation rate for $B_0=2$ T. The low temperature regime is dominated by a first-order process between the states $\ket{\bar{2}}$ and $\ket{1}$. According to Eq. 4 and 5 of the main text, it is composed of a flat initial part followed by a linear increase.  At higher temperatures, the second-order process mediated by phonons between the states $\ket{\bar{2}}$ and $\ket{\bar{3}}$ becomes dominant. We can better understand its functional form by expanding the terms in Eq. 6 of the main text:   

\begin{equation}
\Gamma^{(2)}_{\bar{2}\bar{3}} \propto \int_{0}^{\omega_d} \int_{0}^{\omega_d} \abs{\sum_{k\neq 2,3} \frac{c_{2k} \; c_{k3}}{\Delta E_{\bar{2}k} - \hbar \omega' + \frac{1}{2} i \hbar \Gamma_k}}^2 \omega'^5 \omega''^5 [1+n_b(\hbar \omega')] n_b(\hbar \omega'') \delta(\Delta E_{\bar{3}\bar{2}}+\hbar \omega' - \hbar \omega'') d\omega' d\omega'' ,
\label{eq:ramanRel}
\end{equation}

where $\omega_d$ is the Debye frequency and the coefficients $c_{ij}$ come from the overlap between the states $i, j$ due to mixing between spin and valley. In silicon, the electron-phonon interaction is mediated by deformation potential phonons. Therefore, the matrix elements have an additional factor $\omega ^2$ with respect to the standard interaction with piezoelectric phonons, because of the $\sqrt{q}$ dependence of the strain caused by deformation potential phonons, where $q$ is the wave number. 

As discussed in the main text, $\frac{1}{2} \hbar \Gamma_k$ represents the energy width of the k state, determined by its lifetime. Since the ground state of the system $\ket{1}$ has, at least at low temperature, a long lifetime compared to the state $\ket{4}$, $\Gamma_4 \gg \Gamma_1$, we can neglect the transitions through the state $\ket{4}$ in the sum of Eq.~\ref{eq:ramanRel}. In the following, we will consider separately the contributions to the integral coming from off-resonant ($\hbar \omega' \neq \Delta E_{\bar{2}1}$) and resonant ($\hbar \omega' \approx \Delta E_{\bar{2}1}$) phonons.

In the off-resonant case and at sufficiently high temperatures, phonons with frequencies $\hbar \omega \gg \Delta E_{\bar{2}1}, \Delta E_{\bar{3}\bar{2}}$ are well populated and Eq.~\ref{eq:ramanRel} can be rewritten as:
\begin{equation}
\Gamma^{(2)}_{\bar{2}\bar{3}} = C_{R} T^{9} \int_{0}^{\hbar \omega_d / k_{\textrm{B}}T} \frac{e^x}{(e^x-1)^2}dx 
\end{equation} 
and the relaxation rate scales to a good approximation as $T^9$. In the intermediate regime $\Delta E_{\bar{2}1} \gg \hbar \omega \gg \Delta E_{\bar{3}\bar{2}}$, the term $\hbar \omega'$ in the denominator of Eq.~\ref{eq:ramanRel} can be neglected and the relaxation rate scales as $T^{11}$. In our experimental case, the energy differences between the levels are comparable with each other and thus this last regime is not visible in the experimental data. Instead, if we consider coupling with orbital states, these conditions apply and a $T^{11}$ dependence is expected. The power laws we found are strictly related to the power of the $\omega$ terms in Eq.\ref{eq:ramanRel}, which depends on the particular nature of the electron-phonon interaction. For example, in GaAs, where piezoelectric  phonons dominates over deformation potential phonons, the power is reduced to three instead of five, which leads to a $T^5$ and $T^7$ temperature dependence. In case of Johnson mediated relaxation an even weaker temperature dependence is obtained.  

In the resonant case, we have $\hbar \omega' \approx \Delta E_{\bar{2}1}$ and Eq.~\ref{eq:ramanRel} can be approximated as:
\begin{equation}
\Gamma^{(2)}_{\bar{2}\bar{3}} = C_{O} \frac{[1+n_b(\Delta E_{\bar{2}1})] n_b(\Delta E_{\bar{2}1} + \Delta E_{\bar{2}\bar{3}})}{\Gamma_{1}} ,
\end{equation} 
where the lifetime of the $k$ state is in general evaluated as the inverse of the sum of all first-order processes between $k$ and the other states and it is ultimately limited by the time scale of the experiment. At sufficiently low temperatures, $\Gamma_k$ is temperature independent and the relaxation rate depends exponentially on the temperature according to $\Gamma_{\bar{2}\bar{3}}^{(2)} \propto  n_b(\Delta E_{\bar{2}1} + \Delta E_{\bar{2}\bar{3}})$. At higher temperatures, $\Gamma_k$ becomes also proportional to $n_b(\Delta E_{\bar{2}1} + \Delta E_{\bar{2}\bar{3}})$ and the relaxation rate is given approximately by $1+n_b(\Delta E_{\bar{2}1})$, which is temperature independent for $k_{\textrm{B}}T \ll \Delta E_{\bar{2}1}$ and linear dependent for $k_{\textrm{B}}T \gg \Delta E_{\bar{2}1}$. In our experimental case, this linear dependence is masked by the the Raman process. The resonant and off-resonant transitions can thereby explain all the different regimes that we see in Fig.~\ref{Fig:AllRates} (a).   

The rates in first and second-order are used to solve the 4x4 system of coupled differential rate equations: 

\begin{equation}
\begin{bmatrix}
\dot{N_1} \\
\dot{N_2} \\
\dot{N_3} \\
\dot{N_4} \\
\end{bmatrix}
=
\begin{bmatrix}
-(\Gamma_{12}+\Gamma_{13}+\Gamma_{14}) & \Gamma_{21} & \Gamma_{31} & \Gamma_{41} \\
\Gamma_{12} & -(\Gamma_{21}+\Gamma_{23}+\Gamma_{24}) & \Gamma_{32} & \Gamma_{42} \\
\Gamma_{13} & \Gamma_{23} & -(\Gamma_{31}+\Gamma_{32}+\Gamma_{34}) & \Gamma_{43} \\
\Gamma_{14} & \Gamma_{24} & \Gamma_{34} & -(\Gamma_{41}+\Gamma_{42}+\Gamma_{43})  \\
\end{bmatrix}
\cdot
\begin{bmatrix}
N_1 \\
N_2 \\
N_3 \\
N_4 \\
\end{bmatrix} .
\end{equation}

$N_i$ being the population of the state $i$. For each temperature we extract the four eigenvalues of the matrix. Among the four, one equals zero and corresponds to the stationary population of the levels after the relaxation process is over. Two are much greater than the inverse time scale of the experiment and are therefore discarded, since they correspond to exponential decays not observable in the experiment. Finally, the remaining one represents the time constant that characterizes the single exponential decay of the spin-up fraction as a function of load time. This rate is shown in Fig.~3 (a), (b) and (c). 

As discussed in the main text, the spin lifetime can be further improved by working in a low magnetic field and high valley splitting regime. Fig.~\ref{Fig:AllRates} (d) show this depedence at a temperature of 4 K, where second-order phonon processes dominate the relaxation process. Even at this relatevely high temperature, we extract lifetimes larger than 1 ms for a valley splitting close to 1 meV, which is a very promising result for future scalability of these systems. 

We did not discuss relaxation due to the residual \textsuperscript{29}Si nuclei. However, the presence of nuclei mainly affects the dephasing of the electron spin rather than relaxation, due to the large Zeeman energy mismatch. The modulation of hyperfine coupling by phonons is also suppressed in natural silicon due the low concentration of \textsuperscript{29}Si nuclei \cite{tahan2014relaxation}. The effect can be expected to be even smaller in our case, where the substrate is made of \textsuperscript{28}Si.

\section{Section 3: Measurement of electron temperature and lever arm of the quantum dot}
Both the base electron temperature and the lever arm of the quantum dot have been extracted by a unique measurement, where the width of the charge transition (0$\rightarrow$1) shown by the red arrow in Fig.~1 (b) of the main text is measured as a function of the nominal fridge temperature \cite{prance2012single}. The charge stability diagram shown in Fig.~1 (b), is measured via a double-lockin technique, where the transconductance $dI_s/dP_1$ of the sensing dot is measured by applying an AC excitation $V_{AC}$ to the gate P1. During the map, the current $I_s$ of the sensing dot is kept at the most sensitive point by using a digitally-controlled feedback. The width of the transition is determined by $V_{AC}$ for large AC excitations and by the thermal broadening due to the finite electron temperature $T_e$ when $V_{AC}\ll k_{\textrm{B}}T_e$. In these conditions the transconductance $dI_s/dP_1$ is proportional to the derivative of the Fermi-Dirac distribution:
\begin{equation}
\frac{d I_s}{d P_1} = a \cosh[-2](\frac{\alpha_{P1} (P_1 - b)}{2 k_\textrm{B} T_e}) + c ,
\label{Eq:FermiDirac}
\end{equation}  
where $a$, $b$ and $c$ are fitting parameters and $\alpha_{P1}$ is the lever arm of the quantum dot. The electron temperature $T_e$ depends on the nominal fridge temperature $T_f$ and the base electron temperature $T_0$ according to:
\begin{equation}
T_e = \sqrt{T^2_0 + T^2_f} .
\label{Eq:BaseElecTemp}
\end{equation}
We fix the gate P2 such that the tunneling rate between dot and reservoir is maximized and therefore the signal to noise ratio in the charge stability diagram is also maximized. We  sweep the gate P1 in the direction of the first charge transition. During the sweep we apply an AC excitation to the gate P1 of 15 $\mu$V at 133 Hz.

\begin{figure}[]
	\centering
	\includegraphics[width=0.5\linewidth]{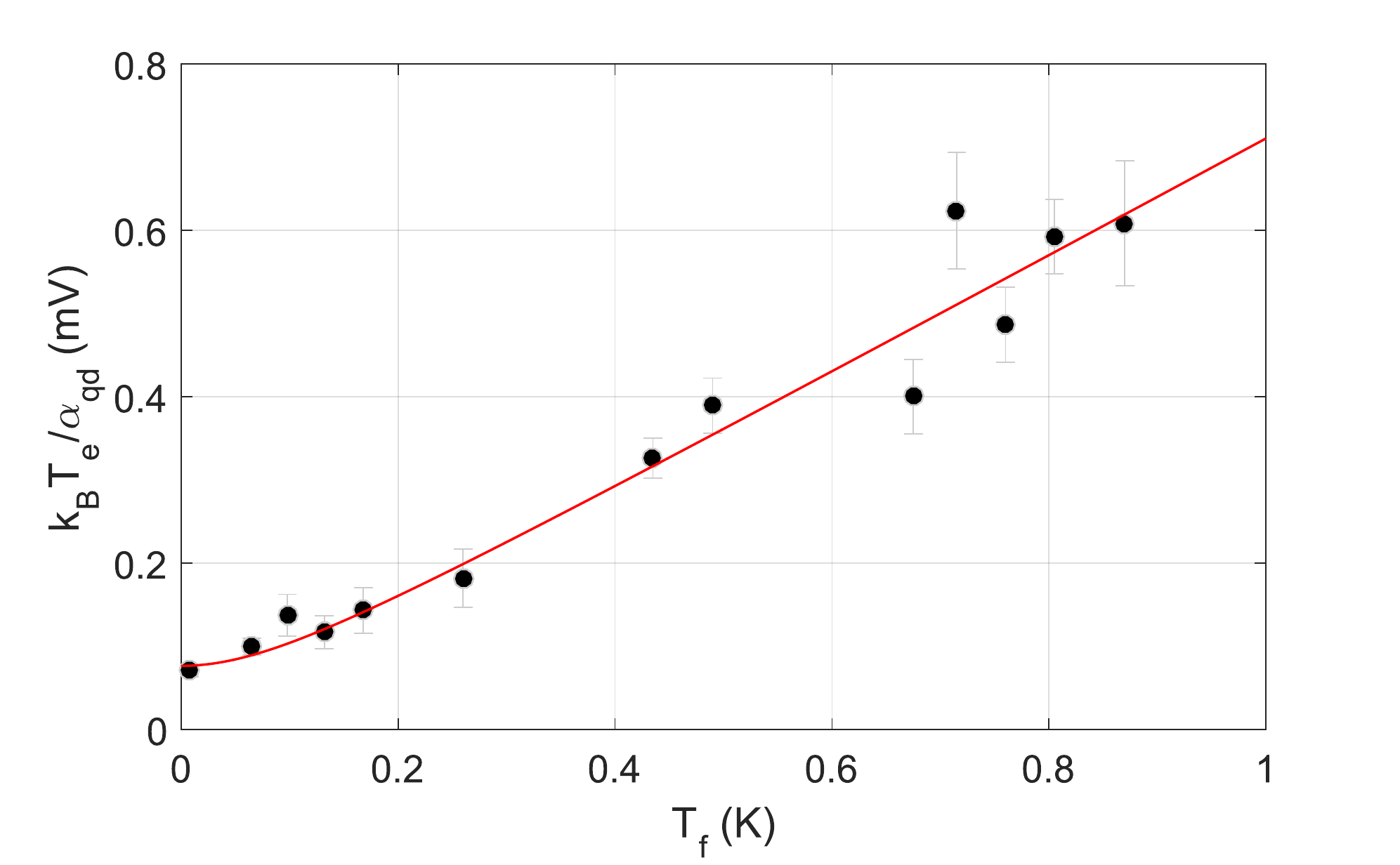}
	\caption{Width of the (0$\rightarrow$1) charge-state transition as a function of the fridge temperature. The red line is a fit according to Eq. \ref{Eq:FermiDirac} and \ref{Eq:BaseElecTemp}. At sufficiently high temperatures, $T_e$ equals $T_f$ and the transition width increases linearly with a slope proportional to the inverse of the lever arm. At low temperature $T_f$ becomes smaller than $T_0$ and the transition width becomes independent of $T_f$. } 
	\label{Fig:ElecTemp}
\end{figure}

Fig.~\ref{Fig:ElecTemp} shows the width of the transition as a function of $T_f$. The width is for all points much higher than the excitation applied to the gate P1 meaning that we are in the conditions of a thermally limited transition. From the fit we extract a lever arm of $\alpha_{P1}=0.122\pm0.005$ eV/V and a base electron temperature of $T_0=108\pm13$ mK. 

\end{document}